# Charge-carrier dynamics in single-wall carbon nanotube bundles: A time-domain study

**Tobias Hertel[1]\*, Roman Fasel[2] and Gunnar Moos[1]**

[1]Fritz-Haber-Institut der Max-Planck-Gesellschaft, Faradayweg 4-6, 14195 Berlin, Germany
[2]EMPA Dübendorf, Überlandstr. 129, 8600 Dübendorf, Switzerland

**Abstract.** We present a real-time investigation of ultrafast carrier dynamics in single-wall carbon nanotube bundles using femtosecond time-resolved photoelectron spectroscopy. The experiments allow to study the processes governing the subpicosecond and the picosecond dynamics of non-equilibrium charge-carriers. On the subpicosecond timescale the dynamics are dominated by ultrafast electron-electron scattering processes which lead to internal thermalization of the laser excited electron gas. We find that quasiparticle lifetimes decrease strongly as a function of their energy up to 2.38 eV above the Fermi-level – the highest energy studied experimentally. The subsequent cooling of the laser heated electron gas down to the lattice temperature by electron-phonon interaction occurs on the picosecond time-scale and allows to determine the electron-phonon mass enhancement parameter λ. The latter is found to be over an order of magnitude smaller if compared, for example, with that of a good conductor such as copper.

Carbon nanotubes (CNTs) and in particular single-wall carbon nanotubes (SWNTs) have attracted broad interest due to their unique electronic structure and properties. Among the most impressive demonstrations of their potential use in novel technologies are the fabrication of field-effect and single-electron transitors from individual or small bundles of SWNTs [1,2,3,4]. Other conceivable applications include the implementation of CNTs as emitters in field-emission displays [5,6], ultrastrong fibers [7,8], chemical sensors [9,10], super-capacitors [11] and more. As molecular field effect transistors they have already been implemented into simple logic devices with outstanding transport characteristics and promising scaling behavior [12]. As molecular wires, carbon nanotubes exhibit exceptional resilience with respect to current induced failure and sustain current densities exceeding $10^9$ A/cm$^2$ before suffering fatal damage, as reported by various groups [13,14,15,16,17,18]. The carrier scattering processes governing linear and non-linear transport properties in these devices, such as electron-electron (*e-e*), electron-phonon (*e-ph*) or impurity scattering are most frequently investigated by conventional – *i.e.* time-independent – transport studies. Here we present a complementary *time-domain* study of carrier dynamics in bundles of single-wall carbon nanotubes using femtosecond time-resolved photoemission. This technique can probe carrier dynamics in real-time and allows to access a wider energy range than typical transport studies.

Time-resolved photoemission was developed from two-photon-photoemission [19,20] and has become a powerful technique for studies of surface and bulk electron dynamics on the femtosecond time-scale [21,22,23,24,25,26,27,28]. It has been most frequently used to probe the relaxation of photoexcited electrons with energies ranging from a few hundred meV above the Fermi level up to the vacuum level. Here, we have extended the most commonly used pump-probe scheme to allow an investigation of charge carriers from 2.38 eV above $E_F$ to the immediate vicinity of the Fermi level and below, down to about -0.2 eV. A detailed analysis of these experiments allows to study fundamental scattering processes such as *e-e* or *e-ph* scattering directly in the time domain.

We find that *e-e* interactions in bundles of SWNTs give rise to strongly increasing *e-e* scattering times as the carrier energy approaches the Fermi level. These processes are relevant not only for non-linear transport phenomena but are also important for a variety of spectroscopies probing electronic properties of these materials. The observed short electron lifetimes ranging from 100 fs at 0.3 eV relative to $E_F$ down to less than 10 fs at energies above 2.38 eV are expected to lead to a significant broadening of spectral features in SWNT bundles. The internal thermalization of the laser excited electron gas is facilitated by these *e-e* interactions and can be characterized by a time constant of 200±50 fs.

\* Corresponding Author

On the other hand we find that *e-ph* interactions are weak and consequently, that the rate of energy transfer between electrons and lattice is slow. The *e-ph* coupling parameter λ obtained from our experiments is found to be 0.006±0.002 which is over an order of magnitude smaller than the coupling constant of a metal like Cu, for example [29]. Such weak coupling of electrons to phonons may contribute to the high resilience of nanotubes to current induced damage.

The paper is organized as follows: in section 1 we will briefly discuss the experimental setup. In section 2 we will present a brief review of the optical properties of SWNT bundles together with an analysis of features found in UV-vis absorption spectra. The principles underlying time-resolved photoelectron spectroscopy and the analysis of our time-resolved photoemission data will be discussed in section 3 before the article ends with a summary and conclusions in section 4.

## 1 Experimental

### 1.1 Sample preparation and UHV system

Single-wall carbon nanotube samples used in this study were made from as-produced soot and from commercial nanotube suspension (tubes@rice, Houston, Texas) with similar results for both types of samples. The commercial nanotube suspension, containing SWNTs with a diameter distribution peaked at 12Å, is used to fabricate bucky paper samples according to the procedure described in reference [30]. The SWNT-paper was attached to a small tantalum disk making use of adhesive forces after wetting both sample and Ta-substrate with a droplet of ethanol. Slow drying and the use of thin paper (less than 0.5 mg cm⁻²) were required to prevent the sample from peeling off the substrate. The sample temperature was measured using a type K-thermocouple attached to the tantalum disk. As a reference we mounted a sample of highly oriented pyrolytic graphite (HOPG) on the backside of the Ta-disk, using silver paint to facilitate good adhesive and thermal contact. The HOPG sample was cleaved directly before being transferred into the vacuum chamber.

A schematic illustration of the sample holder is shown in Fig. 1. The sample could be heated resistively by a pair of tantalum wires on which the Ta-disk was suspended. The temperature across the sample was homogeneous to within less than one degree as estimated using the high temperature edge of thermal desorption spectra from rare gas monolayers adsorbed on the HOPG surface. The bucky paper samples were outgassed thoroughly by repeated heating and annealing cycles under ultra high vacuum conditions with peak temperatures of 1200 K.

The sample holder was attached to a He-continuous flow cryostat (Cryovac) which allowed sample cooling down to about 40 K. The position of the sample within the chamber was controlled with an *xyz*-manipulator equipped with a differentially pumped rotary feed-through to allow control of the azimutal angle. Ultra high vacuum (UHV) of typically 1·10⁻¹⁰ mbar was maintained by a combination of a membrane, turbo-drag and turbo-molecular pump (Balzers).

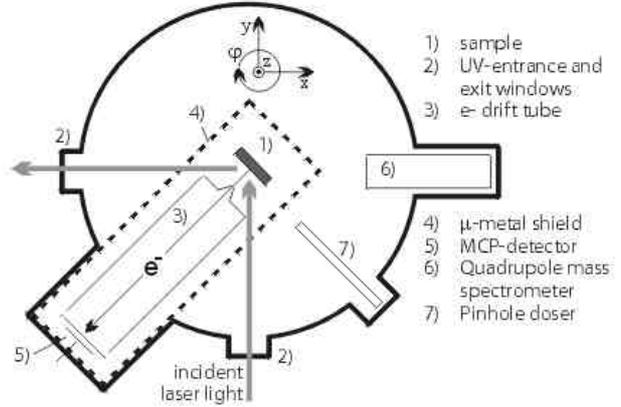

a) UHV-chamber

1) sample
2) UV-entrance and exit windows
3) e- drift tube
4) μ-metal shield
5) MCP-detector
6) Quadrupole mass spectrometer
7) Pinhole doser

incident laser light

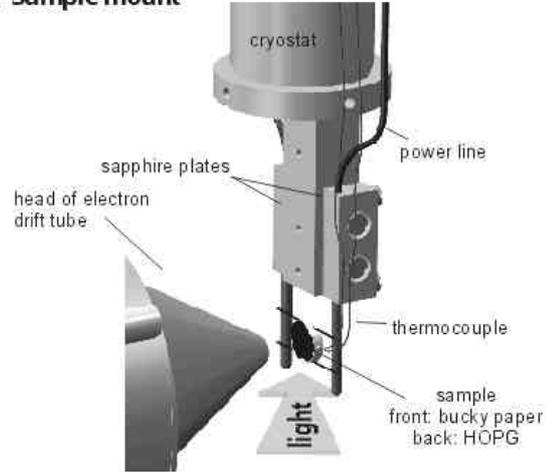

b) Sample mount

cryostat

sapphire plates

power line

head of electron drift tube

thermocouple

sample
front: bucky paper
back: HOPG

light

**Fig. 1**. Schematic illustration of the experimental setup. a) UHV chamber and some of its components. b) Sample mount.

All experiments were performed on a number of samples and on different spots of each sample to ensure reproducibility of the data.

### 1.2 Laser system and optics for time-resolved spectroscopy

Femtosecond laser pulses at 810 nm are generated by a Ti:sapphire oscillator (Coherent MIRA 900) and subsequently amplified to a fluence of 4 mJ/pulse with 200-kHz repetition rate by a regenerative amplifier (Coherent RegA 9000). The RegA output pumps an optical parametric amplifier (Coherent OPA 9400) to generate tunable femtosecond pulses with wavelengths between 470 nm and 730 nm and with a pulse energy of typically 100 nJ. These are compressed using a pair of SF10 prisms to nearly Fourier-limited pulses of 50–85 fs at full width at half maximum. The compressed OPA output is focused into a 0.2-mm-thick BBO crystal (type I) by a f = 200 mm lens, generating the second harmonic beam. The fundamental and second harmonic beams are separated by a di-chroic mirror and guided to the UHV chamber. The time delay between visible pump and UV probe pulses is adjusted by a computer-controlled delay stage. The beams are focused onto the sample non-

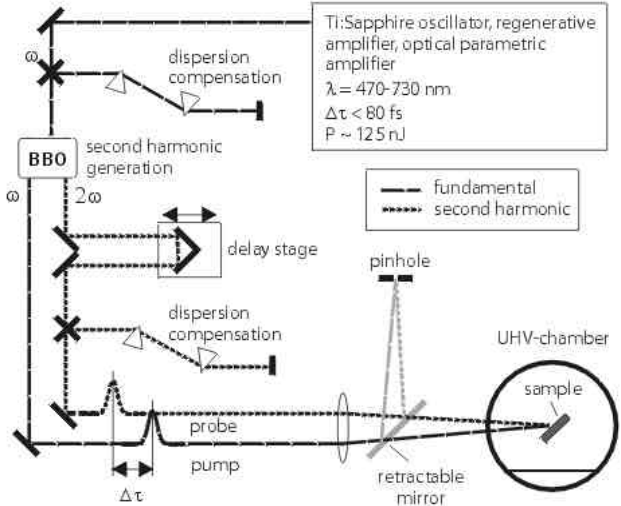

Laser system and fs-optics

**Fig. 2.** Schematic illustration of the optical set up for time-resolved photo-electron spectroscopy.

collinearly (skew 2°) by a fused silica lens positioned outside the UHV chamber ($f = 300$ mm). Both beams enter the UHV chamber through a 1-mm-thick $MgF_2$ window and are incident on the sample at an angle of 45° with respect to the surface normal. The resulting fluence for the visible pump and UV probe beams is typically 50 $\mu J/cm^2$ and 5 $\mu J/cm^2$, respectively. This corresponds to a pump power of ~ $10^9$ $W/cm^2$. The spatial overlap is adjusted by aligning the two beams outside the UHV chamber using a pinhole of 150 $\mu$m diameter. The pulse cross-correlation width is determined from the cross-correlation signal of higher energy photo-electrons from polycrystalline copper or tantalum in combination with the known (ultrashort) lifetimes of excited electrons in these materials [24,31]. A schematic illustration of this setup is shown in Fig. 2. For more details see reference [32].

### 1.3 Time-of-flight photoelectron detection

Photoemitted electrons are detected by a time-of-flight (ToF) spectrometer, which is shielded electrically and magnetically by a cylindrical $\mu$-metal (Co-Netic) tube of 1.5 mm thickness. After entering a graphitized drift tube through a 3 mm diameter orifice, electrons drift through a 30 cm long field-free region and are accelerated onto a pair of multi-channel plates after exiting the drift tube through a highly transparent copper grid. The electron time-of-flight distributions are acquired by a PC using the output of a time-to-amplitude converter. The signal of a fast photo-diode from the white light leakage of the OPA is used as start signal for the time-to-amplitude conversion while the amplified output of the MCP detector is fed through constant fraction discriminator and then used as stop signal. This results in an overall time resolution of approximately 250 ps. The energy resolution of the spectrometer was 10 meV at 1 eV kinetic energy. The angle of acceptance of the ToF spectrometer assembly is approximately 100 mrad.

## 2 Optical properties of SWNT bundles

In this section we will present a brief review of the electronic structure of SWNT bundles and the resulting optical properties. We will also discuss a scheme that may allow to obtain semi-quantitative information on the abundance of different tube types and the corresponding distribution in the chiral vectormap. The discussion in this section is helpful to understand the character of the optical excitations in SWNT bundles by which we generate the non-equilibrium distribution and initiate charge carrier dynamics.

The optical properties of SWNT samples [33,34] have previously been studied by a number of groups using ultraviolet to visible (UV-vis) absorption spectroscopy as well as electron energy loss spectroscopy.

The electronic structure of SWNTs is closely related and can be derived from that of graphene in the general manner by zone folding the electronic structure of 2D graphite onto the 1D Brillouin zone (BZ) of nanotubes [35]. The way in which the electronic structure of graphene is zone-folded is generally characterized by the chiral (or equatorial) vector of the tube $C_h = na_1 + ma_2$, where $a_1$ and $a_2$ are the unit vectors of the graphene lattice and $(n,m)$ refers to the chiral index of the nanotube. Within the tight binding approximation one finds that all tubes with $n-m=3N$, with $N=0,1,2...$, are metallic while the remainder is semi-conducting with a band gap that scales inversely proportional to the tube diameter $d = |a_1|/\boldsymbol{p}\,(n^2+m^2+nm)^{1/2}$, with $|a_1| = 2.46$Å. For an isotropic distribution of chiral angles $\tan(\theta) = \sqrt{3}\,m/(2n+m)$ one would thus expect that 1/3 of all tubes are metallic and 2/3 are semi-conducting.

For graphite, optical excitation in the visible and UV range up to about 5 eV is dominated by $\boldsymbol{p}$ or $\boldsymbol{p}^*$ inter-band transitions with initial and final states nearly symmetrically distributed above and below the Fermi level [36]. Other electronic transitions, such as $\boldsymbol{s}^* \leftarrow \boldsymbol{p}$ or $\boldsymbol{p}^* \leftarrow \boldsymbol{s}$ play no significant role at the photon energy of ~ 2.4 eV used for optical excitation in this experiment. Note, that Drude absorption contributes increasingly only for photon energies below about 1 eV and the onset of $\boldsymbol{p}$-plasmon excitation is beyond 5 eV [34,36].

As stated above, the electronic structure of SWNTs and consequently their optical properties are closely related to those of graphene. Due to the one-dimensional nature of the electronic bands in SWNTs, however, the density of states (DOS) of SWNTs exhibits a series of characteristic van Hove singularities (VHS) as seen in the lower part of Fig. 3 where we have plotted the density of states for a (9,9) SWNT. The latter has a diameter of 12 Å which is typical for the tubes found in these samples. In the upper part of Fig. 3. we have reproduced the diameter distribution for these samples as obtained from a TEM study [30] along with the corresponding distribution of chiral vectors in the $x$-$y$ plane of graphite (upper right). For simplicity we have assumed that chiral vectors are distributed isotropically. The average DOS, resulting from a superposition of the DOS of all possibly tube types – weighted by their abundance and assuming an isotropic chirality distribution – is plotted in the lower part of Fig. 3. At energies close to $E_F$ one can clearly distinguish between clusters of VHS arising from semi-conducting and metallic nanotubes. The first two clusters – associated with semi-conducting tubes – are found to be centered around $0.12\gamma_0$ and $0.24\gamma_0$ while the cluster(s) around $0.36\gamma_0$ originate from VHSs of metallic nanotubes. Here $\gamma_0$ is the nearest neighbor hopping matrix element with reported values ranging from 2.5 to 3.0 eV [37] depending on the experimental technique used for its determination (here we used $\gamma_0$=2.65 eV).

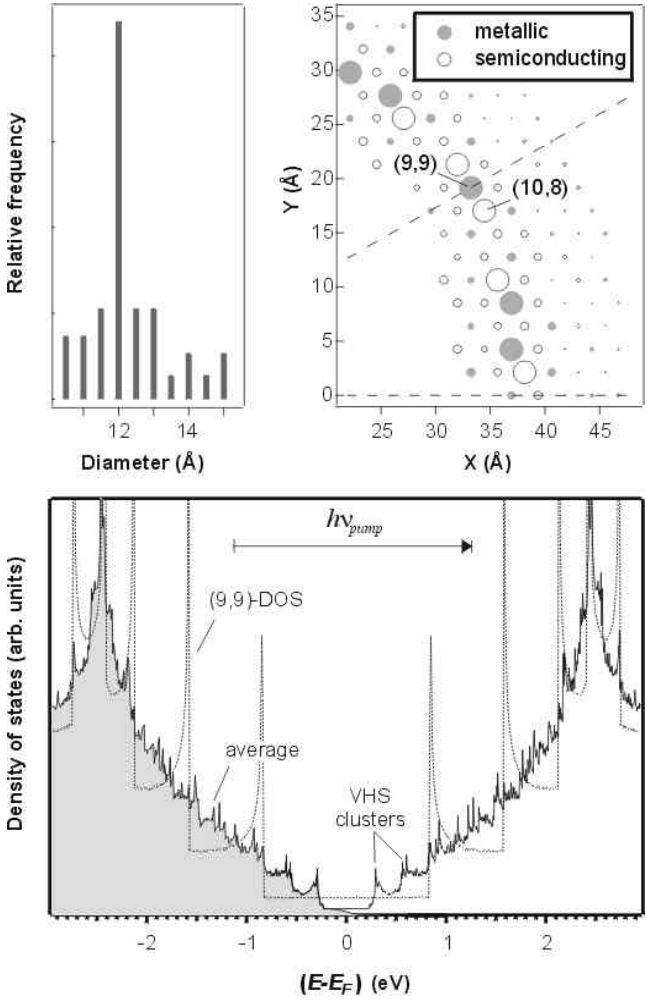

**Fig. 3.** Upper left panel: diameter distribution of the material used in this study as published in ref [30]. Upper right panel: chiral vector map for an isotropic distribution with diameters according to the experimentally determined frequencies (the statistical weight is proportional to the area of the markers). Lower panel: calculated average DOS. The dashed line is the DOS of the (9,9) tube, one of the most probable species in these samples.

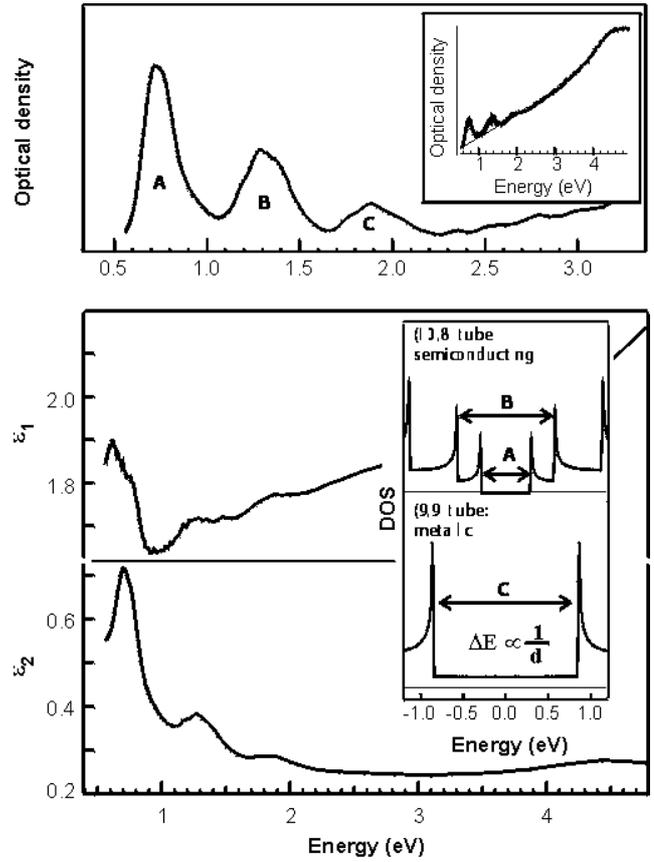

**Fig. 4.** Background corrected UV-vis absorption spectrum from SWNT bundles (upper panel). The inset shows the full spectrum from the un-annealed sample before background subtraction. The three most prominent features labeled A,B and C correspond to clusters of optical transitions between different subsets of van Hove singularities (VHS). A and B are due to transitions between the first VHS in semi-conducting tubes and C is associated with transitions between the first VHS in metallic tubes. The lower part shows the dielectric functions as computed from the measured absorption, transmission and reflection spectra.

The average DOS suggests that similar structures associated with clusters of VHS may also be found in the optical spectra and dielectric function of SWNT bundles. It can be inferred from Fig. 4 – where we have reproduced optical spectra of a thin (un-annealed) SWNT film – that this is indeed the case. Here one also observes three pronounced features in the spectra which can tentatively be assigned to symmetrical transitions between the first two clusters of VHS's of semi-conducting and transitions between the first cluster of metallic SWNTs, respectively.

The dielectric functions reproduced in Fig. 4 where determined in the usual manner from an independent measurement of the transmission $T(w)$, absorption $A(w)$ and reflection $R(w)$ coefficients. For the calculation of the dielectric functions we have accounted for multiple scat-tering within the films. Absolute values of the dielectric functions – in particular that of $\varepsilon_2$ – are somewhat uncertain due to difficulties in determining the exact film thickness. In this analysis the latter was treated as a free parameter to satisfy $T+A+R=1$ among these independently measured quantities.

The optical constants obtained in this manner lie somewhere in-between those of graphite and $C_{60}$ crystals and are in reasonable agreement with results from an EELS study by Pichler *et al.* [34,36]. Using the dielectric function reported in Ref. [34] we obtain an optical skin depth at 2.38 eV of about 50 nm. This is relatively close to the value of ~30 nm obtained for optical excitation with polarization perpendicular to the c-axis in graphite. The skin depth calculated from the dielectric functions shown in Fig. 4 would be larger by nearly a factor of 10. This discrepancy may partly be due to the uncertainty in the determination of $\varepsilon_2$. The latter value is actually closer to the skin depth of about 0.5 μm obtained for optical excitation with light polarized parallel to the c-axis in graphite.

A closer look at the absorption spectra in Fig. 4. reveals that the clusters also contain some fine structure which appears to consist of a series of evenly spaced features. This becomes even more evident in systematic studies with sam-ples of different diameter distributions by Jost *et al.* [39] and Kataura *et al.* [33]. A random distribution of tube types and chiralities – such as the one used for illustrative purposes in Fig. 3 – however, is expected to lead to more and smaller spaced features in the absorption spectra [39]. This discrepancy was taken as further evidence for anisotropic grouping of chiral angles around 30° along the armchair axis [38] which also appears to agree with Raman and electron diffraction studies [39,40].

A more detailed analysis, however, requires a calculation of the optical properties of SWNT bundles. To this end we have used a somewhat simplified approach by calculating the optical response of SWNT bundles within the tight binding approximation for light polarized parallel (∥) to the tube/bundle axis. This simplifies the calculations greatly because the selection rules for excitation with $E_\parallel$ specify that the only allowed transitions are those with $\Delta J=0$, *i.e.* symmetric transitions with respect to $E_F$. $J$ is the angular momentum belonging to the wave-functions of different sub-bands [41,42]. Excitation by $E_\perp$, which requires that $\Delta J=\pm1$, leads to absorption features in-between $A$ and $B$ in Fig. 4 and cannot give rise to features in the energy region from $B$ to $C$ [41] which stretches from 1.1 eV to 2.6 eV and will be analyzed in the following. In this case, a restriction to optical transitions with $E_\parallel$ thus seems appropriate. A more detailed discussion which also includes the excitation by light polarized perpendicular to the tubes can be found, for example in ref. [41].

The distinct advantage of the above approximations is that for excitation with $E_\parallel$ we need to consider only symmetrical transitions between $\pi$ and $\pi^*$ sub-bands. The corresponding joint density of states (JDOS) can easily be obtained from the DOS by simply rescaling the energy axis by a factor of two. This is appropriate within the tight binding approximation due to symmetrical $p$ and $p^*$ bands. The JDOS is then used in the computation of optical spectra as approximation for the optical absorption function. Fortunately, the JDOS reproduces all distinct absorption features obtained from a full calculation of the dielectric functions [41]. The use of the JDOS instead of the optical absorption function basically amounts to neglecting the energy dependence of transition matrix elements.

In the following, we determine the best agreement of experimental spectra with the JDOS by numerically averaging over a certain distribution of $(n,m)$ indices as characterized by a small set of free parameters. We do not attempt to describe spectra quantitatively but rather aim at a qualitative simulation which focuses on the position and width of absorption features and also tries to account for some of the fine structure. As shown below, the agreement between calculated and experimental spectra is striking. The trial-distribution that was used for the superposition of individual JDOS's is characterized by a set of parameters ($q$, $\Delta q$, $d$, $\Delta d$), where $q$ and $\Delta q$ are the mean chiral angle of the Gaussian chiral-angle distribution of width $\Delta q$ and $d$ as well as $\Delta d$ are the corresponding parameters for the mean and the width of the Gaussian diameter distribution.

The spectra where broadened by a Gaussian of 20 meV FWHM. The data was initially fit under certain constraints, which were removed later on. In a first step we used a trial distribution with small chiral spread of 6° around the mean chiral angle in the zigzag ($q=0°$) or armchair ($q=30°$) direc-tions and allowed the other parameters – including the tight binding parameter $\gamma_0$ – to vary freely. The average diameter $d$ was constrained to values within ±1Å of the mean diameter reported in reference [30]. The results of a best fit to the 'zigzag-' and 'armchair-distribution' are reproduced in the upper and lower panels of Fig. 5, respectively. From the upper curves it is evident that the observed spectra can hardly be described by a zigzag distribution of finite width. If the chiral spread is allowed to vary at a fixed mean a chiral angle of 0°, we find best agreement for a nearly isotropic distribution (see middle panel of Fig. 5). The same procedure for the armchair distribution gives best agreement with the experiment for a chiral spread $\Delta q$ of 6°. Even though the results of this fit not necessarily yield the *only* possible distribution that can account for the observed spectra, this provides further evidence for previous conclusions that the distribution of chiral angles appears to be grouped around the armchair direction [38,39,40]. The procedure described here should become a powerful tool for a quantitative analysis of SWNT samples if combined with a thorough calculation of optical constants and accurate diameter statistics as obtained from diffraction studies, for example.

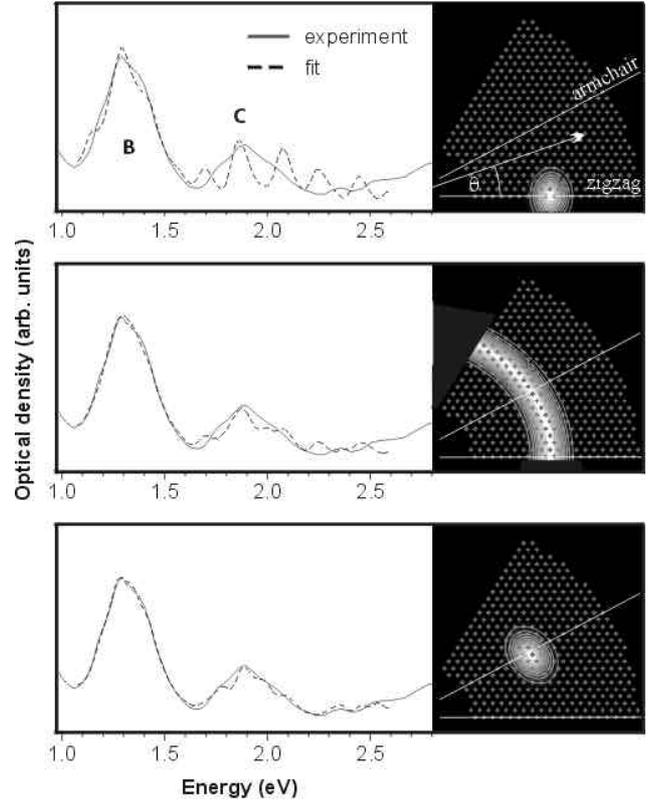

**Fig. 5.** Comparison of experimental and calculated spectra as obtained from a superposition of the JDOS of different tube types. The distribution used for the averaging process is shown on top of the chiral vector map on the right. Constraints used in the fit where: mean chiral angle ($q=30°$) and chiral spread ($\Delta q=6°$) (upper panel), isotropic chiral spread (middle panel) and mean chiral angle ($q=30°$) (lower panel). The best agreement with experimental spectra is obtained for the armchair distribution in the lower panel, which also corresponds to the global optimum for this type of chirality distribution.

A few additional details about the results of this fit procedure seem to be worth mentioning. First: best agreement with experimental spectra was obtained a) if the width of the diameter distribution was allowed to be larger than reported in reference [30] with a spread about equal to that obtained in ref. [38] and b) for a tight binding parameter $\gamma_0$ of $2.65\pm0.01$ eV. Comparable but significantly poorer agreement was only found for $\gamma_0$=2.91 eV, in which case the mean diameter had to be extended to 13.5 Å which is about 1.5Å higher than the value measured for these samples by TEM [30] and thus seems too high. Allowing the mean chiral angle to differ from $\mathbf{q}$=30° did not lead to any significant improvement of the fit. We furthermore note, that spectra cannot be fit by a combination of armchair tubes only. These result in features with a spacing $\Delta d$=0.07 nm which corresponds to some of the fine structure in the cluster $C$ but the 'armchair-only' distribution would entirely fail to account for the shape of the $A$ and $B$ features.

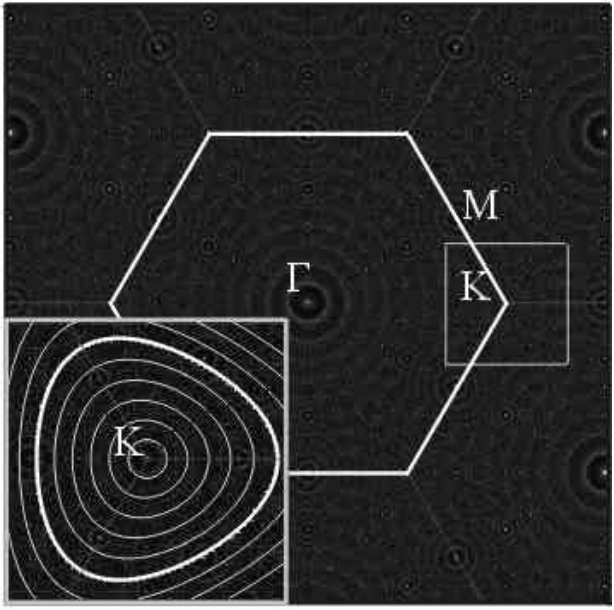

**Fig. 6.** Calculated density of states of SWNT bundles in reciprocal space if projected on the graphene Brillouin zone. Bright areas indicate regions with high density of states. The inset shows the region around the K-point of the Brillouin zone on an expanded scale. The thick solid line in the inset corresponds to regions in k-space in which excitation by 2.4 eV photon is expected to create non-equilibrium carriers.

The coupling between different tube types within one and the same bundle and the resulting changes of the optical properties may also lead to some changes in these spectra and should be studied in more detail experimentally as well as theoretically.

A last point of interest in this section is the average DOS in reciprocal space. We define this average $\mathbf{k}$-space DOS using:

$$G(\mathbf{k}) \propto \sum_{n,m} \sum_{\mathbf{m}} f_{n,m} \int \left(\nabla_\mathbf{k} E_\mathbf{m}\right)^{-1} \mathbf{d}\left(\mathbf{k} - \mathbf{k}_{n,m}^{\mathbf{m}}(E)\right) dE \quad (1)$$

where the $f_{n,m}$ is the statistical weight of a certain tube species specified by its index $(n,m)$ and $\mu$ is an index which runs over all bands of a specific tube type. In Fig. 6 we have calculated this average $\mathbf{k}$-space DOS if projected on the 2D Brillouin zone of graphene before zone folding. For illustrative purposes we have again averaged over the chirality vector map shown in Fig. 3. This distribution also clearly shows the clustering of states around high symmetry points in the BZ. The areas of interest for our experiments are regions in which photons of 2.3 eV – 2.4 eV energy can create excited carriers. For $\Delta J$=0 transitions this corresponds to the vicinity of the thick solid line in the inset of Fig. 6.

The average DOS $D(E)$ reproduced in Fig. 3 can now also be obtained from $G(\mathbf{k})$ using:

$$D(E) \propto \sum_\mathbf{m} \int_{BZ} E_\mathbf{m}(\mathbf{k}) G(\mathbf{k}) d\mathbf{k} \quad (1)$$

where the integral is over the first graphite Brillouin zone (BZ).

Following the arguments of the above discussion, we thus conclude that optical excitation by 2.38 eV photons will lead to nearly equally strong excitation of semi-conducting and metallic tubes by $\pi$ to $\pi^*$ interband transitions, with carriers excited predominantly to energies in the vicinity of $\pm\frac{1}{2}h\mathbf{n}_{\text{pump}}$ with respect to $E_F$. In reciprocal space this is associated with excitation to states predominantly around the K-points in the graphene BZ as shown in Fig. 6. The actual carrier distribution, however, will almost immediately deviate from this idealized picture as scattering processes rapidly redistribute energy and momentum within the system. The latter processes will be discussed in the following.

## 3 Results and discussion of time-resolved photo-emission experiments

In this section we continue with a detailed analysis and discussion of experimental results from the time-domain measurements with particular focus on the influence of $e$-$e$ and $e$-$ph$ interactions on carrier dynamics in SWNT bundles.

### 3.1 Principle of time-resolved photoelectron spectroscopy

Two-photon photoelectron spectroscopy can be used to investigate dynamical processes on the pico- or femto-second time-scales by utilizing pump-probe techniques which are appropriate for the measurement of ultrafast dynamical processes. In the study presented here, the role of the first photon is to stimulate carrier dynamics in the sample by generating a non-equilibrium carrier distribution through optical excitation (pump) and the role of the second photon is to monitor the resulting electron dynamics by photoemission (probe). The photon energies typically used for such studies are in the visible and the UV range for the pump and probe beams, respectively.

The information provided by a time-resolved photo-emission study is to some respect complementary to that obtained from conventional transport studies. The experiments presented here provide us with very detailed information on energy relaxation in contrast to conventional transport studies. The strength of the latter lies in their sensitivity to momentum relaxation which may allow to extract phase and momentum relaxation lengths/times for carrier energies within a small energy window with $|E-E_F|<k_BT$. Their sensitivity to energy relaxation and dynamical processes at higher energies, however, is rather limited.

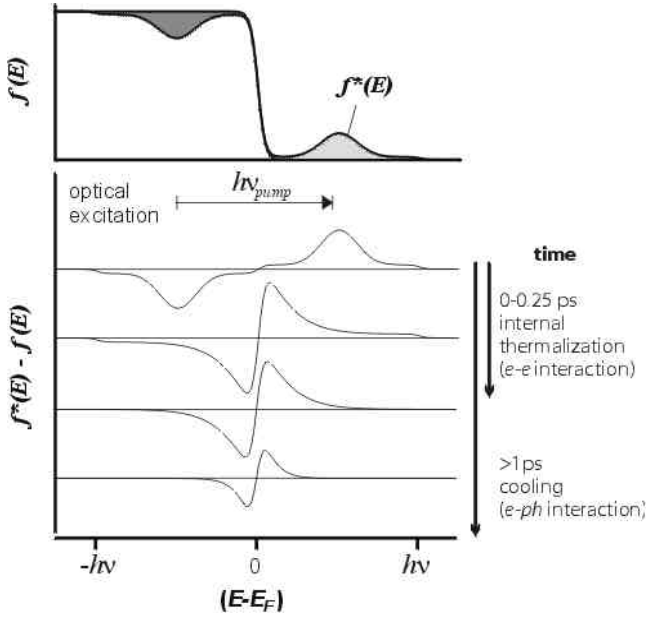

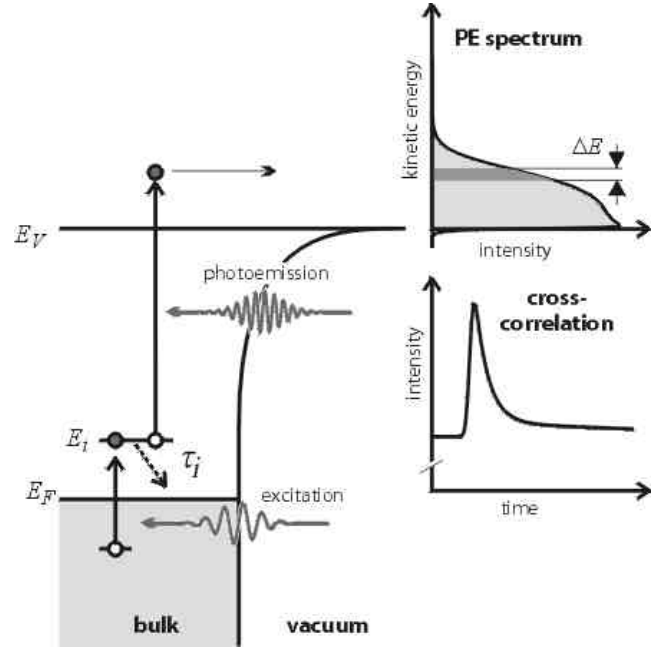

**Fig. 7.** Schematic illustration of the non-equilibrium carrier dynamics as induced and probed optically in a time-resolved photoemission experiment. The initial non-equilibrium carrier distribution is produced by absorption of a visible laser pulse. Electron-electron interactions lead to internal thermalization of the non-equilibrium carrier distribution on the sub-picosecond timescale. Electron-phonon interactions in turn are responsible for a cooling of the laser heated electron gas down to the lattice temperature on the picosecond timescale.

**Fig. 8.** Illustration of the pump-probe scheme employed in time-resolved photoemission. A visible laser pulse excites a non-equilibrium electron distribution. A second UV laser pulse then probes the state of the electronic system by photoemitting carriers into the vacuum. Measurement of the photoelectron intensity within a certain energy interval as a function of time delay between pump and probe pulses then allows to determine the decay time $t_i$ associated with relaxation of the electron population in the intermediate states $E_i$.

The present investigation is thus complementary to classical transport studies by providing information on carrier dynamics in a broader energy range directly in the time-domain.

The dynamical evolution of a non-equilibrium carrier distribution initiated by light absorption is illustrated schematically in Fig. 7. The change of the distribution function resulting from optical absorption is indicated by the first curve in the lower part of Fig 7. As discussed in the previous section, excitation of tubes with their axis parallel to the polarized light ($\Delta J$=0) will lead to an excitation profile nearly symmetrical with respect to $E_F$ while the excitation of those lying perpendicular to the electrical field ($\Delta J$=±1) will lead to some excitation at slightly higher and lower energies ($\frac{1}{2}h\nu_{pump}$ ± 0.12 $\gamma_0$ ≈ 1.7 eV ± 0.3 eV) and thus effectively smear out the excitation profile.

The dynamics of the resulting non-equilibrium carrier distribution $f^*(E)$ can be described in the general manner by a Boltzmann equation:

$$\frac{df^*}{dt} = \left(\frac{\partial f^*}{\partial t}\right)_{e-e} + \left(\frac{\partial f^*}{\partial t}\right)_{e-ph} \quad (3)$$

where the first and second terms on the right refer to the changes of the distribution function due to *e-e* and *e-ph* interactions, respectively.

In the following we will discuss qualitative aspects of the evolution of such a non-equilibrium carrier distribution and later on follow up with a detailed analysis of the experiments.

The non-equilibrium distribution generated by the absorption of visible light will regain thermal equilibrium by a combination of *e-e* and *e-ph* interactions. Electron-electron interactions, for example, are most efficient in redistributing the energy of the excited electrons among all charge carriers in the system and thereby lead to what is generally referred to as *internal thermalization* of the electronic system. This redistribution of energy proceeds through rapid generation of a growing number of secondary *e-h* pairs which are created in a cascading manner by each *e-e* or hole-hole (*h-h*) scattering event. The generation of two additional non-equilibrium carriers in each such scattering process rapidly leads to the formation of a secondary electron cascade. This will eventually lead to a new equilibrium of the electronic system at somewhat higher temperature after which *e-e* interactions do not lead to any further changes of the distribution function. Note that electron-hole re-combination is too slow and does not play a significant role for the dynamics on the picosecond and sub-picosecond time-scales. Electron-phonon interactions on the other hand transfer energy from the electronic system to the lattice and are responsible for a cooling of the electron gas back to the lattice temperature on the picosecond time-scale.

The major challenge in the analysis of time-resolved photoemission data lies in the task to disentangle the different contributions to the non-equilibrium carrier dynamics arising from *e-e* and *e-ph* interactions respectively. Fortuitously, we find that the characteristic time-scales on which *e-e* and *e-ph* interactions modify the carrier distribution in SWNT bundles, are well separated and thus easy to identify.

Experimentally, the dynamics are probed using the time dependence of the photoemission signal after perturbation of the system by a visible pump pulse. This process is illus-trated schematically in Fig 8. The intensity of the photo-emission signal at a particular kinetic energy can then be used to study the decay of carriers in the intermediate state $E_i$ by varying the time-delay between the pump and probe pulses. The resulting, so called, cross-correlation trace (XC) then directly reflects the dynamics in the intermediate state and thus the time scale $\tau_i$ on which the population changes.

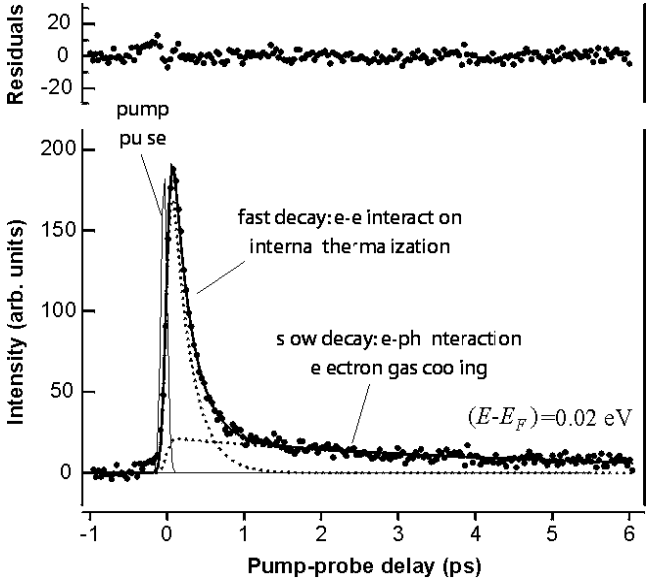

**Fig. 9**. Sample fit of a low energy cross correlation trace (XC) to a bi-exponential decay. The slow and fast components can be associated with *e-e* and *e-ph* interactions, i.e. internal thermalization and cooling of the electron gas, respectively.

A typical XC and bi-exponential fit to the data is reproduced in Fig. 9. The relaxation of charge carriers from such low energy XCs ($E$-$E_F$ = 20 meV) is found to occur on two distinct time-scales. As we will discuss in further detail below, these two components can be associated with internal thermalization and electron gas cooling, *i.e.* with *e-e* and *e-ph* interactions, respectively.

### 3.2 Photoelectron spectra from SWNT bundles

The photon energies used for probing photoelectron dynamics in the SWNT sample and the HOPG reference were chosen to exceed the sample work function by typically 100-300 meV. In doing so we were able not only to record the electron dynamics in normally unoccupied states above the Fermi level but also the dynamics directly above and below $E_F$. (see Fig. 10) The initial state energy ($E$ - $E_F$) of photoemitted electrons with respect to $E_F$ is obtained from their kinetic energy $E_{kin}$ using ($E$ - $E_F$) = $E_{kin}$ + $h\nu_{probe}$ - $e\Phi$ where $e\Phi$ is the sample work function. The work function of these samples is determined from the width of photoelectron spectra and was found to be 4.52±0.05 eV.

The variations of $e\Phi$ when scanning across the sample surface where significantly smaller than 50 meV. This is also apparent from the extraordinary sharp secondary edge in the photoelectron spectra which average over a 150 µm diameter area probed by the UV laser beam.

The change in the photoelectron signal after perturbation by the visible pump pulse is seen most clearly in the difference spectrum in the lower panel of Fig. 10. In the following these difference spectra ($\Delta I = I$(excited)-$I$(equilibrium)) will be referred to as 'excitation spectra' because they reflect the changes in the electron distribution function $f(E)$ induced by excitation of the system with the pump pulse. The intensity of electrons originating from below the Fermi level, for example, is seen to decrease due to excitation of carriers to states above $E_F$ and the intensity of electrons above the Fermi level is thus found to increase (this leads to the negative and positive intensities in the excitation spectrum). Note, that fluctuations of the intensity below the Fermi level are due to noise in the PE signal. The position of the low-energy cut-off depends on the amount by which the probe photon energy exceeds the sample work function. The high-energy cutoff on the other hand can best be seen in the excitation spectrum on a logarithmic intensity scale and is determined by the pump-photon energy (see inset Fig. 10).

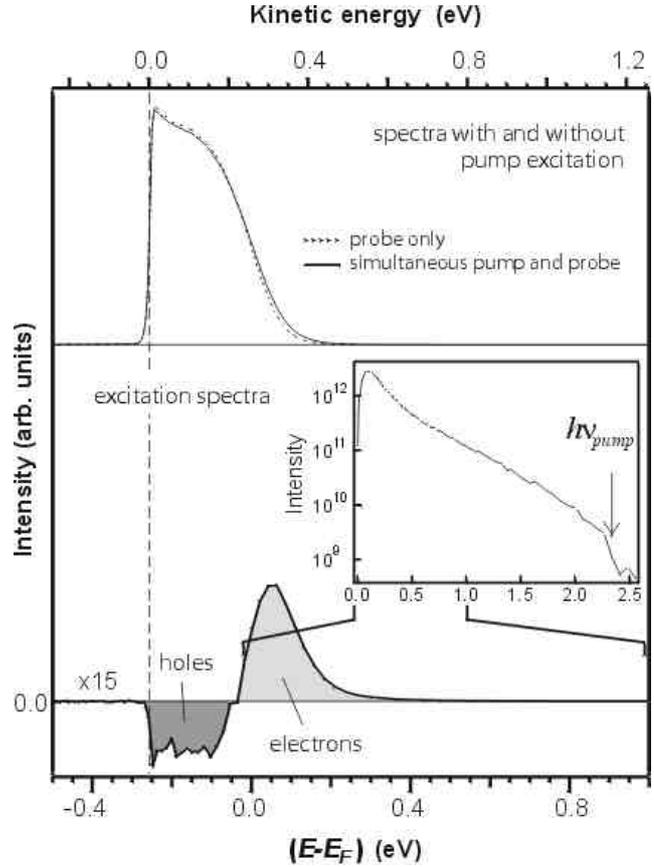

**Fig. 10**. Photoelectron spectra from SWNT bundles: Before (dashed line) and after (solid line) perturbation by a visible pump pulse ($h\nu_{pump}$ = 4.76 eV, $h\nu_{probe}$ = 2.38 eV). The strong increase of the signal on the low energy side of the spectrum is due to photoemission from the Fermi level. The changes of the photoelectron distribution after pump excitation are magnified in the lower panel. The scatter in the lower spectrum increases strongly for small kinetic energies due to the larger background from direct photoemission at these energies. The inset with the excitation spectrum on a log scale shows that only few carriers are excited up to the energy of the pump pulse.

In contrast to UV-vis spectra, however, one does not find any features in excitation spectra that can be associated with clusters of VHS (see Fig. 4, for example). This can be attributed to the high sensitivity of photoemission to the alignment of spectral features with respect to the Fermi level and is indicative of residual doping and small shifts of the band structure of individual nanotubes. UV-vis spectroscopy on the other hand is rather insensitive to this alignment since it primarily probes energy differences between bands.

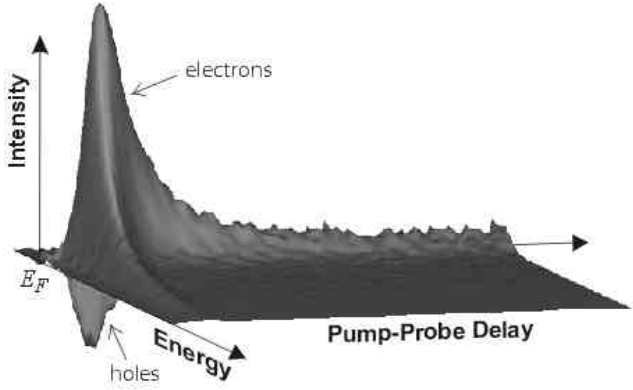

**Fig. 11.** Two-dimensional map of real-time charge-carrier dynamics in single-wall carbon nanotube bundles after perturbation of the electronic system by a visible pump pulse. The intensity shows how the electron and hole populations are generated and evolve in time. The energy axis is 2 eV wide and the pump-probe delay ranges from –0.3 ps to 2.5 ps. The photon energies were 2.32 eV and 4.64 eV for the pump and probe beams, respectively.

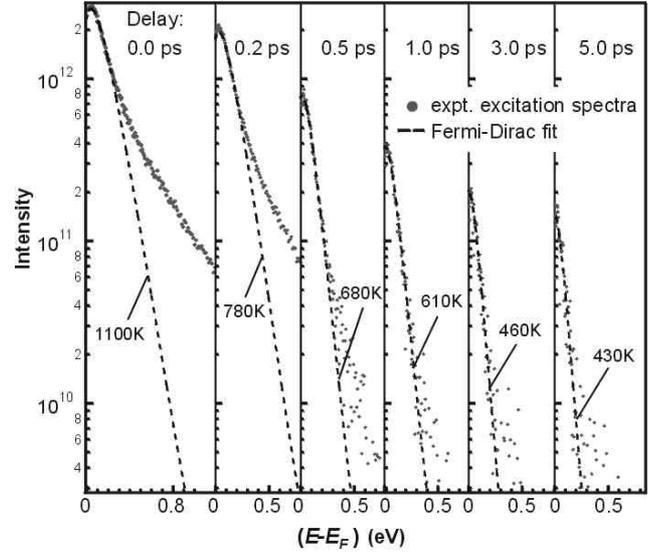

**Fig. 12.** Spectral dynamics in SWNT bundles after excitation by a visible pump pulse. The deviations of experimental excitation spectra from a Fermi-Dirac fit clearly show how the initial non-equilibrium carrier distribution first approaches a new internal equilibrium at elevated temperature and then cools down to the lattice temperature ($T_l$=300 K).

Interestingly, excitation spectra – such as the one in Fig 10 – decay nearly exponentially up to the high energy cutoff at $h\overline{n}_{pump}$. From the discussion in the previous section one might expect to see a pronounced maximum in these spectra around ½ $h\overline{n}_{pump}$ due to the dominance of symmetrical $\pi$ to $\pi^*$ interband transitions during the excitation process. We note in passing that similar spectra are found for excitation spectra in graphite. The absence of spectral features at energies to which carriers are expected to be excited by the pump pulse, indicates that dynamical processes lead to substantial changes of the electron distribution already on the time-scale given by the length of the exciting laser pulse of 85 fs.

These dynamical changes are most clearly seen if we plot the excitation spectra as a function of the time delay between the visible pump and UV probe pulses (see Fig. 11). This map of the electron dynamics clearly reveals how carriers are initially excited and then decay on the subpicosecond and picosecond time-scales.

### 3.3 Spectral analysis

The dynamics of excited charge carriers can be studied either by analyzing the photoelectron signal within a certain energy interval or by investigating the spectral shape as a function of pump-probe time-delay. Here we will begin with a discussion of the evolution of excitation spectra for different time-delays from 0-5 ps (see Fig. 12). Evidently the electron distribution can not be described by a fit to a FD-distribution at all times. Such a fit to FD statistics is frequently used to determine sample temperatures from the width of the Fermi edge in photoelectron spectra. Here it is utilized to reveal deviations of the electron distribution from an equilibrium distribution and to get a rough idea of the electronic temperature, *i.e.* the internal energy of the electronic system. The FD-fit is performed by fitting the product of the sample DOS and the FD distribution $f$ to the experimental intensities $\Delta I$. We also accounted for the fact that excitation spectra reflect the changes of the electron distribution induced by the visible pump pulse by subtraction of the FD distribution prior to excitation with temperature $T_l$, where $T_l$ is the lattice temperature. The resulting expression used for the FD fit was:

$$\Delta I\left(E,T_e(t)\right) \propto D(E)\left(f\left(E,T_e(t)\right) - f\left(E,T_l\right)\right) \quad (4)$$

where the electronic temperature $T_e(t)$ at a certain pump-probe time delay $t$ and a proportionality constant were the only free parameters.

The experimentally derived distribution is found to deviate from the FD fit in particular at small pump-probe delay and higher electron energies because the electronic system requires some small but finite time to regain its internal equilibrium after perturbation by the visible pump pulse. The time-scale of 200±50 fs for this internal thermalization can be estimated from the decrease of the linear deviation of the FD fit from the experimental data. The scattering processes responsible for the formation of a new equilibrium are those which allow a fast redistribution of the energy initially introduced into the system by optical excitation. In metallic or semi-metallic systems *e-e* scattering processes are generally most efficient in doing so. For simple metals, Landaus theory of Fermi-liquids predicts that – on the average – half of the excess energy of a non-equilibrium charge carrier ½($E$-$E_F$) is transferred to a secondary electron-hole pair excitation in a single scattering event. The generation of secondary excitations sets off a whole series of scattering events which rapidly redistribute the excess energy of the primary excitation among a growing number of non-equilibrium charge carriers. This eventually leads to internal thermalization of the electronic system – generally on the subpicosecond time-scale in agreement with the observations reported here.

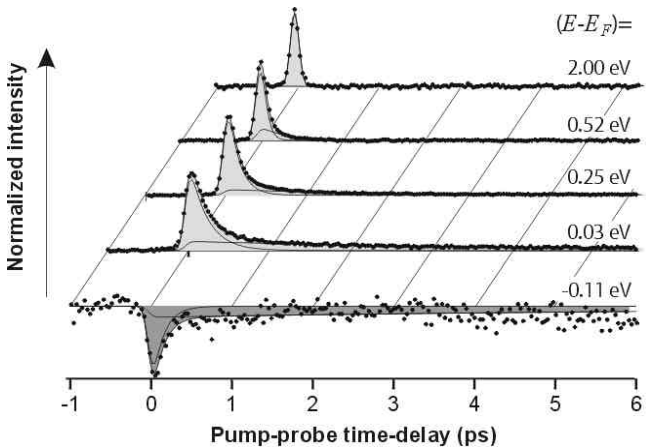

**Fig. 13.** Waterfall plot of normalized cross-correlations. The energy interval for which these cross correlations were recorded varies from –0.11 eV (lowest trace) up to 2.0 eV. The dynamics can be well described by a bi-exponential decay.

The initial rise of temperatures in the FD fit to over 1000 K indicates that the electronic system is heated substantially by the visible pump pulse. The observed temperature jump roughly agrees with expectations if we calculate the temperature rise assuming that the electronic heat capacity $c_e$ of the tube sample is linear in $T_e$ and similar to that of graphite with a heat capacity coefficient **$g$** of 2.4 J/m$^3$/K$^2$ [43]. The linear temperature dependence of the electronic heat capacity was also verified experimentally by measuring the peak temperature rise at different pulse fluences. This revealed that the peak temperature scales approximately with the square root of the pulse fluence as expected for a linear temperature dependence of $c_e$.

For longer delay times the temperature of the FD fit is found to decrease slowly and approach the lattice temperature. This reflects the coupling of electrons and lattice by $e$-$ph$ scattering which eventually leads to a full accommodation of the two systems on the picosecond time-scale.

Since the time-scales for internal thermalization and cooling of the electronic system are well separated we can discuss the two processes on an individual footing. In the following we will first discuss the process of internal thermalization and then continue with a more detailed discussion of the cooling dynamics and electron-lattice interactions.

### 3.4 Internal thermalization and e-e interactions

In the last section we stated that $e$-$e$ scattering processes most likely dominate the internal thermalization dynamics of the laser excited electron gas in nanotube bundles. Strong evidence for this can be obtained from a detailed analysis of cross-correlation (XC) traces. A small set of normalized XC-traces is shown in Fig. 13 for electron/hole energies ranging from –0.11 eV up to 2.00 eV. Evidently, the decay of the electron/hole populations after perturbation by the visible pump pulse is strongly energy dependent. At the highest energies we observe a simple exponential decay of the electron population with a characteristic decay-time close to the limit of our experimental time-resolution of about 10 fs. For energies not too close to the Fermi level this decay time can be associated with the electron or hole quasiparticle (QP) lifetime in the material. At energies closer to $E_F$, XCs are increasingly influenced by secondary electron cascades, $i.e.$ by a refilling of states by electrons decaying from states higher in energy, as well as by competition with $e$-$ph$ interactions. As a consequence, the decay of the electron population closer to $E_F$ does not directly reflect QP lifetimes. The contribution of secondary electron cascades to the measured decay can be calculated using detailed rate equations for electron dynamics as discussed by Gusev and Wright [44]. This shows that the decay of the electron population at energies ($E$-$E_F$) above about 0.3 eV is dominated by direct QP decay and the interpretation of XCs is straightforward.

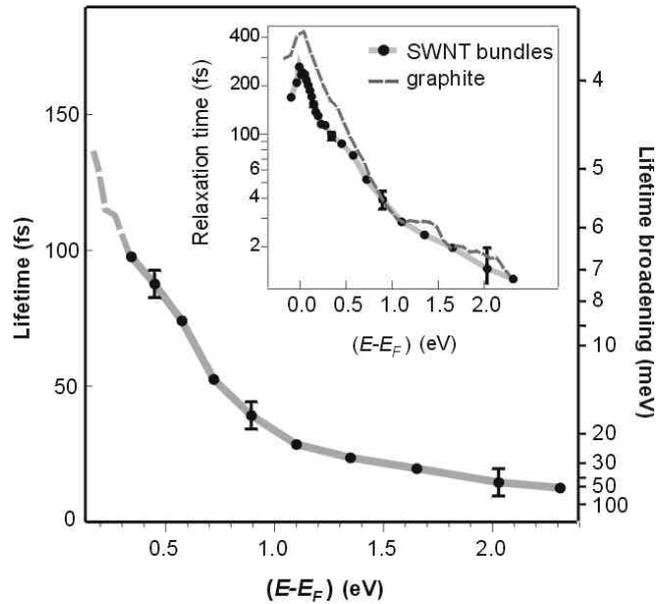

**Fig. 14.** Energy dependence of quasiparticle lifetimes as obtained from the fast component of the bi-exponential fit to cross-correlation traces. At energies close to the Fermi level the observed relaxation times are increasingly influenced by electron-phonon interactions as well as by secondary electrons and cannot be directly associated with quasiparticle lifetimes. The inset shows relaxation times over the entire energy range probed in these experiments in comparison with results obtained for graphite [45].

For energies approaching the Fermi level we observe an additional slow decay-channel contributing to the XCs. This slow process reflects the decrease of the electron population due to cooling of the entire electronic system which can be inferred from the decrease of the electronic temperature in Fig. 12. Due to the presence of this slow channel we use a bi-exponential fit to analyze XCs and to extract the energy dependence of the fast process that is associated with QP lifetimes.

The decay times of the fast-component obtained from the bi-exponential fit are plotted in Fig. 14 as a function of the carrier energy. The graph clearly shows a strong energy dependence of the electron relaxation with decay times rising from below 10 fs at 2.3 eV up to 280 fs near $E_F$. In metals, such a pronounced energy dependence of lifetimes is characteristic of energy relaxation by $e$-$e$ scattering. In graphite, $e$-$e$ interactions were also found to be the dominant energy relaxation channel through a comparison with recent $ab$ $initio$ self energy calculations [45,46]. The similarity of the results presented here with those obtained for graphite in a similar study (see dashed line in Fig. 14) suggests that *e-e* scattering processes are likewise responsible for the fast decay in SWNT bundles.

To schematically illustrate the possible consequences of finite QP lifetimes for the electronic structure of SWNTs we have convoluted the tight-binding density of states of a prominent tube type, the semiconducting (10,8) tube, with an energy dependent lifetime (see Fig. 15). The latter was obtained from the fit of an empirical trial function to the QP lifetimes in Fig. 14 [47]. We have also plotted the computed lifetime broadening $\Delta E$ on the right axis of Fig. 14, with $\Delta E$(FWHM)=0.65 eV fs/$t$. The results of this convolution in Fig. 15 clearly illustrate how spectral features such as van Hove singularities in the DOS may be broadened due to *e-e* interactions, in particular at higher energies. It should be noted, however, that the photoemission signal averages over contributions from all different tube types in these samples and, therefore, is not necessarily representative for the carrier lifetime in a particular state of a particular tube type.

A somewhat stronger broadening of van Hove singularities has been found in scanning tunneling spectroscopic (STS) studies of individual dispersed SWNTs or small bundles of SWNTs [48,49]. Smaller broadening, on the other hand, was found in recent experiments by Jorio *et al.* who determined the width of VHSs using resonant Raman scattering from individual SWNTs [50]. The latter authors report on very narrow features in the energy dependence of the resonant Raman signal with a broadening of the order of 0.1 meV to 1 meV. The lifetime broadening predicted for this energy range by our time-domain measurements is somewhat larger with about 10 meV. These differences between the apparent linewidths in different experiments may be attributed to the different environments in which the tubes in each study are situated as well as to the method by which the linewidths are determined. The largest widths are apparently found in STS studies where tubes are commonly dispersed on metallic noble metal substrates. Unless excessively broadened by resonant tunneling into the metal, the intrinsic STS linewidths should be comparable to or smaller than quasiparticle lifetimes in the underlying substrate. QP lifetimes for the noble metals copper and silver are rather similar to the ones observed in this time-domain study, while lifetimes of excited carriers in transition metals with a higher density of states near $E_F$, like tantalum for example, are generally found to be significantly smaller [51]. However, in addition to the coupling to bulk excitations or resonant tunneling into the substrate other mechanisms may also contribute to broadening of STS linewidths and make a direct comparison with results from this time-domain study difficult. The resonant Raman experiments on the other hand, were performed on individual SWNTs and on insulating substrates where the coupling to substrate excitations is expected to be negligible. In the study presented here, we observe smaller lifetimes – corresponding to larger linewidths – than in the Raman experiments. This can either be attributed to a stronger coupling to secondary excitations in SWNT bundles (if compared to individual tubes) but the apparent discrepancy may also be a consequence of the previously mentioned averaging process over contributions from different tube types in photoemission. In the following we will present arguments as to why the resonant Raman experiments yield a much smaller linewidth for a particular VHS and how these results may be explained by QP decay due to the *e-e* interactions.

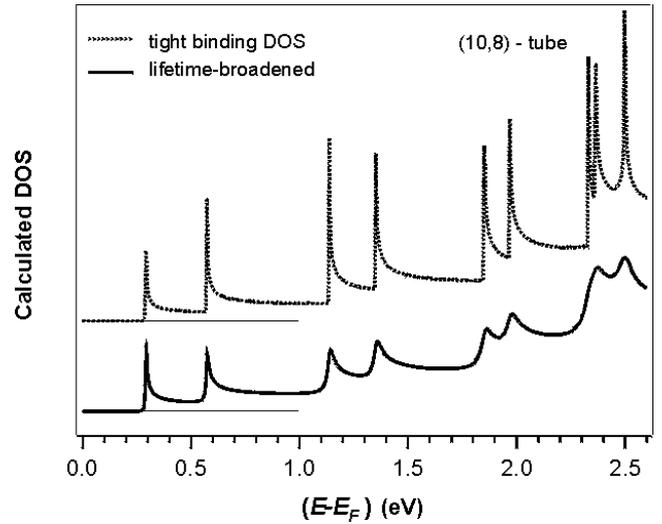

**Fig. 15**. Lifetime-broadened DOS for a semi-conducting nanotube. The tight binding DOS was convoluted with an energy dependent lifetime as obtained from the fast decay of photo-excited carriers. Note, however, that the average lifetimes reproduced in Fig. 14 are not necessarily representative of the lifetimes in a particular tube type like this (10,8) tube and that some features may in fact be broadened more and some less severely. This graph is merely intended to illustrate how finite electron lifetimes may lead to a broadening of van Hove singularities.

The decay of excited carriers in simple metals such as copper, for example, has been studied in much detail experimentally as well as theoretically and is now well understood [24,25,52,53]. However, only recently have experimental and theoretical studies revealed the influence of the peculiar band-structure of graphite on its quasiparticle lifetimes [45,46]. One of the key factors that determine the energy dependence of QP lifetimes in all these systems is the phase-space volume available for *e-e* scattering processes. Within Landau's theory of Fermi-liquids, the energy dependence of the latter leads to the famous $(E-E_F)^{-2}$ dependence of QP lifetimes in free electron systems. In graphite the phase-space volume furthermore depends strongly on the direction of the QP momentum. Lifetimes in graphite have been reported to vary by over an order of magnitude for iso-energetic electrons from different points in the Brillouin zone which could – at least qualitatively – be explained using phase space arguments [45,46]. In contrast to this, one finds that QP-lifetimes in noble metals – with their comparatively isotropic band-structure – show a much weaker directional dependence on $k$.

The concept of the phase space volume has thus been extremely helpful in developing a detailed understanding of the decay of excited carriers in a variety of systems. In the following we will extend such arguments to predict some of the qualitative features of QP lifetimes in SWNTs. This discussion somewhat simplifies the problem of QP decay in SWNTs but should still be appropriate to foresee some of the general trends and possibly point to some interesting details which are overcast by the averaging of contributions from different tube species in photoemission.

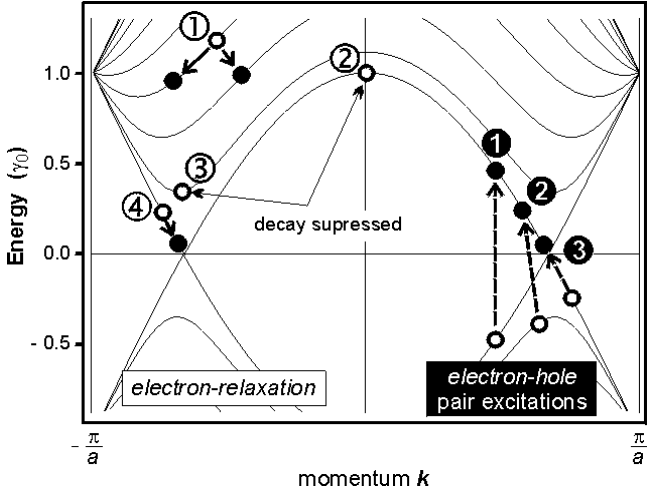

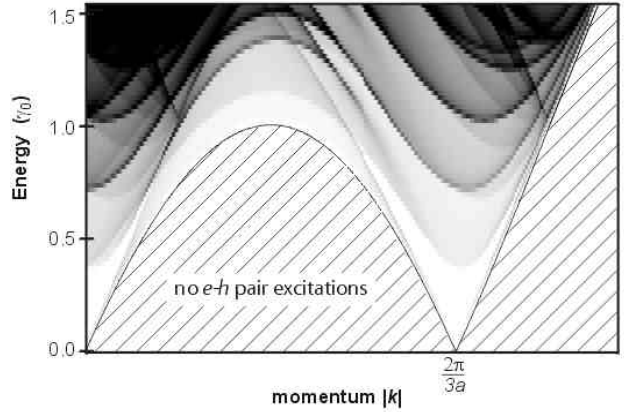

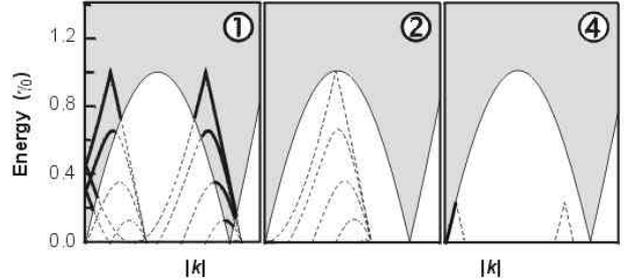

**Fig. 16.** Schematic illustration of different electron decay processes and electron-hole (*e-h*) pair excitations in a metallic (9,9) SWNT. Decay rates not only depend strongly on the carrier energy but also on the initial state sub-band due to constraints imposed on *e-e* scattering processes by momentum and energy conservation.

The relaxation of excited carriers by *e-e* interactions is generally accompanied by the generation of *e-h* pairs to which energy and momentum of the decaying electron are transferred. The probability, *i.e.* the rate, of QP decay thus directly depends not only on the final states available to the decaying electron but also on the number of possible *e-h* pair excitations that can accommodate the corresponding energy and momentum change. Some possible decay processes and *e-h* pair excitations are illustrated schematically for a metallic (9,9) tube in Fig. 16. The complete *e-h* pair spectrum of this tube is shown in Fig. 17a. The 'density' P of *e-h* pair excitations of energy $e$ and momentum $|\mathbf{k}|$ is given by the integral:

$$P(e, |\mathbf{k}|) \propto \iint d\mathbf{k}_h\, d\mathbf{k}_e\, d(\mathbf{k}_h + \mathbf{k}_e - |\mathbf{k}|)$$
$$\times d(E(\mathbf{k}_h) + E(\mathbf{k}_e) - e) \qquad (5)$$

For metallic (n,n) tubes this *e-h* pair spectrum has the same basic structure as that of a single graphene layer [47].

The constraints imposed on *e-e* scattering by energy and momentum conservation can be illustrated graphically using the overlap of the *e-h* pair spectrum calculated from eq. (5) with the electron energy-loss spectrum $L(E, \mathbf{k})$ for decay from a particular initial state $|E_i, \mathbf{k}_i\rangle$. We define this loss function as:

$$L(E, \mathbf{k}) \propto \sum_f \int_{E_f \geq E_F}^{E_f < E_i} dE\, \left( \nabla_{\mathbf{k}} E_f\big(|\mathbf{k} - \mathbf{k}_i|\big) \right)^{-1}$$
$$\times d\big(E_i - E_f\big(|\mathbf{k} - \mathbf{k}_i|\big) - E\big) \qquad (6)$$

where the index *f* runs over all final state sub-bands of the particular tube-type under consideration and the integration is performed over final states with energies between $E_F$ and $E_i$. Some of these electron energy loss spectra for decay from different initial states are shown in Fig. 17b. They thus give possible combinations of energy and momentum-change if an excited electron decays from the initial state

$|E_i, \mathbf{k}_i\rangle$. In the following we will discuss the decay from 4 distinct initial states indicated by points 1-4 in Fig 16.

A carrier in the initial state labeled (1) in Fig. 16 can decay by either *inter-* or *intraband* relaxation. *Intraband* relaxation is associated with no change of angular momentum $J$ and thus can only couple to *e-h* pairs which have zero angular momentum, *i.e.* intraband or symmetrical *e-h* pair excitations. Two possible excitations of this kind labeled by filled circles (1) and (3) are shown in Fig. 16. The *interband* decay from state (1) on the other hand is associated with a change of angular momentum $\Delta J = 1$ and thus couples to *e-h* pairs with $J=1$, such as the *e-h* pair labeled (2).

The situation for decay from the initial state (2) is fundamentally different as the electron energy loss spectrum of this carrier has no overlap with the *e-h* pair continuum at all (see middle panel of Fig. 17b). Carriers at the top of the metallic bands in this tube are thus expected to be comparatively long lived. Spataru *et al.* have found that QP decay from the M-point of the graphite Brillouin zone is also suppressed due to similar constraints by momentum and energy conservation. Note, that zone folding transfers the M-point of the graphene BZ to the position indicated by (2) in Fig. 16.

Decay from the initial state (3) is likewise suppressed. In this case however, it is conservation of angular momentum which causes the suppression. The latter is critical because carriers can decay from (3) only via excitation of low-energy e-h pairs with angular momentum J=1. The lowest energy e-h pairs with J=1 are the ones corresponding to the type labeled (2). The threshold for excitation of such e-h pairs, however, is too high to be accounted for by an electron decaying from (3) which effectively leads to a suppression of the decay from the bottom of the first subband. This situation is thus rather similar to that encountered for carriers at the bottom of the conduction band in semiconducting tubes. We speculate that the small width (0.1-1 meV) of the resonance leading to an enhanced Raman signal in the study by Jorio et al. [50] can be attributed to a similar suppression of e-e scattering processes as that described above. As a consequence, we also expect the width of the 'JDOS-resonance' in Raman scattering experiments to be strongly temperature dependent.

**Fig. 17.** Schematic illustration of the constraints by momentum and energy conservation on *e-e* scattering processes. a) *e-h* pair spectrum for non-interacting electrons in a (9,9) tube where *a* is the graphene lattice constant. b) comparison of the *e-h* pair spectrum with energy-loss spectra for an electron decaying from different initial states labeled (1), (2) and (4) in Fig 16. These graphs already allow a qualitative discussion of expected QP lifetimes. The loss spectrum of electrons decaying from the state labeled (2) for example, has no overlap with the *e-h* pair continuum and decay by *e-e* scattering is thus expected to be suppressed.

The last e-e scattering process considered here is that of a carrier with its initial state close to $E_F$ in one of the metallic bands. Decay from state (4) can occur exclusively via intraband relaxation and thus only couples to e-h pairs of the type labeled (3) in Fig. 16. The overlap of the electron energy loss spectrum with such e-h pairs becomes very small for initial state energies approaching $E_F$ and will thus lead to a strong increase of e-e scattering times for electron energies approaching the Fermi level.

The constraints by momentum and energy conservation are thus expected to lead to a strong dependence of QP lifetimes on band index and QP momentum. Similar arguments can be reiterated for other tube types. Characteristic for semi-conducting tubes should be the slow decay of electrons from the bottom of the conduction band. For semiconductors like Si, for example, interband relaxation across the bandgap via phonon emission is slow with decay times typically on the ns timescale [54]. Here we do not observe any decay which might be attributed to interband recombination in semiconducting tubes. This suggests that tube-tube interactions may also facilitate intertube scattering, between semiconducting and metallic tubes, for example. This would open new decay channels for electrons at the bottom of the conduction band of semiconducting tubes and explain the absence of such slow decay processes in the time-resolved photoemission spectra. Note, that nonlinear transport studies in the high-bias regime of multi-wall carbon nanotubes also show that current is carried by more than one shell which is indicative of stronger intershell coupling at higher bias, i.e. at higher electron energies [55]. Such intershell or intertube scattering may also be facilitated by the coupling of excited carriers to low energy collective excitations. In SWNT bundles such collective excitations are facilitated by tube-tube interactions in a similar manner as they are facilitated in graphite through layer-layer interactions.

Disorder may provide an additional source of momentum whereby the constrains for energy and momentum relaxation are somewhat relaxed. For graphite we found that dynamic and static disorder, e.g. phonons or defects, lead to an overall increase of electron relaxation rates [45]. This should also be kept in mind since SWNT samples are generally expected to be structurally less perfect than pristine HOPG on which the experiments of ref. [45] were performed.

We speculate, that future studies on the influence of collective excitations as well as on the influence of dynamic and static disorder on the decay of excited carriers may thus also provide a means to study tube-tube interactions.

### 3.5 Cooling of the electron gas and e-ph interactions

The e-e scattering processes described in the last section lead to a rapid redistribution of energy absorbed by carriers on the sub-picosecond time-scale. After perturbation by the visible pump pulse this brings the electronic system close to internal equilibrium at a somewhat elevated temperature (see Fig. 12). The complete equilibration of electrons and phonons, naturally involves e-ph interactions which transfer energy from the electronic system to the lattice. It can be inferred from Fig. 12 that internal- and electron-lattice thermalization take place on distinctly different time-scales and the processes can thus be analyzed on an individual footing. In this section we will concentrate on e-ph interactions and the resulting cooling of the laser heated electron gas.

The non-equilibrium dynamics of an electronic system coupled to a phonon bath is frequently described in the framework of the two-temperature model [56]. Within this model, the energy flow and the corresponding changes in electron and lattice temperatures, $T_e$ and $T_l$, are described by a set of coupled differential equations:

$$C_e \frac{dT_e}{dt} = \nabla(\boldsymbol{k}\nabla T_e) - H(T_e, T_l) + S(t) \qquad (7)$$

$$C_l \frac{dT_l}{dt} = H(T_e, T_l) \qquad (8)$$

where $C_e$ and $C_l$ are the corresponding heat capacities, $\boldsymbol{k}$ is the electronic heat diffusion coefficient, $S(t)$ a source term given by the pump laser pulse and $H(T_e, T_l)$ is the crucial electron to lattice energy transfer rate. The first term in eq. (7) describes diffusive transport due to temperature gradients in the electronic system. With an electronic c-axis diffusion constant in graphite – as a reference material – of below 0.05 W/cm K these processes are expected to be too slow to be relevant at the time-scale on which we observe electronic cooling. In particular the morphology of the bucky-paper samples is expected to be responsible for an even smaller heat diffusion coefficient perpendicular to the surface. This and the comparatively large optical skin depth of over 50 nm, allows to neglect diffusive transport out of the detection volume and we can omit the carrier diffusion term of eq. (7).

The discussion can be further simplified due to the small electronic heat capacity $C_e$ if compared to that of the lattice $C_l$. The electronic heat capacity of electrons in SWNT bundles can be estimated using the density of states at $E_F$ for a typical metallic tube, e.g. the (9,9) species, of 0.015 atom$^{-1}$ eV$^{-1}$. In passing, we note that this is almost exactly three times higher than the DOS in graphite of $5.5 \cdot 10^{-3}$ atom$^{-1}$ eV$^{-1}$ [57]. By this fortuitous coincidence, one obtains an average specific heat capacity in bundles with a 1:2 mixture of metallic to semi-conducting tubes of approximately 12 µJ/mole K$^2$ which is nearly equal to that of graphite of 13.8 µJ/mole K$^2$ [58]. At room temperature the electronic heat capacity in SWNT bundles is thus expected to be over 4 orders of magnitude smaller than the lattice heat capacity of $7.5 \cdot 10^3$ mJ/mole K [59,60]. The temperature of the lattice thus does not increase noticeably and the differential equations (7) and (8) can be approximated by a single one:

$$C_e \frac{dT_e}{dt} = -H(T_e, T_l) \qquad (9)$$

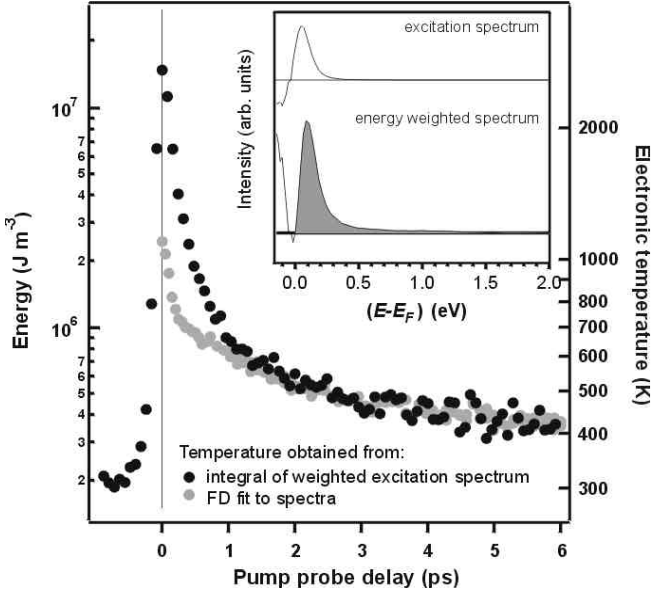

**Fig. 18.** Internal energy of the electronic system as a function of time after excitation by a visible pump pulse. The energy can either be obtained from a FD fit to the excitation spectrum or by integration of the energy weighted spectrum.

In the following we will discuss how the energy transfer rate $H(T_e, T_l)$ can be obtained from time resolved photoemission experiments. To this end we recall that the coupling function $H(T_e, T_l)$ can also be written in terms of the change of internal energy $E_{int}$ of the electronic system:

$$H(T_e, T_l) = -\frac{dE_{int}}{dt} \qquad . (10)$$

The latter can be obtained by integration using:

$$E_{int} = \frac{\int\limits_{-\infty}^{\infty} dE\, E\, D(E)\, f^*(E)}{\int\limits_{-\infty}^{\infty} dE\, D(E)\, f^*(E)} \qquad (11)$$

where $D(E)$ is the density of states in the system and $f^*(E)$ is the appropriate distribution function of the carriers. Note, that $f^*(E)$ doesn't have to be in thermal equilibrium. Experimentally this integral can be obtained from an integration over the energy weighted photoelectron intensity $I(E_{kin})$:

$$E_{int} \propto \int\limits_{-\infty}^{\infty} dE\, E\, I(E + h\boldsymbol{n}_{probe} - e\Phi) \qquad (12)$$

where we assume that the effective photoemission matrix elements are constant – at least within the small energy range with significant spectral weight. The expression in brackets simply relates the measured electron kinetic energy, to the energy with respect to $E_F$ prior to photo-emission by the UV probe photon with energy $h\boldsymbol{n}_{probe}$. $e\Phi$ is the sample work-function. Since the excitation spectra cover only a limited energy range and are cut off at the low energy side by the secondary edge, we calculate the internal energy from photoemission spectra using positive energies only:

$$E_{int} = E_{int}(T_l) + 2\hat{C} \int\limits_{0}^{\infty} dE\, E\, \Delta I(E + h\boldsymbol{n}_{2hn} - e\Phi) \quad (13)$$

where $E_{int}(T_l)$ is the internal energy of the system before optical perturbation while it is still in equilibrium with the lattice. $\Delta I$ is the measured change in the photoemission intensity after the optical perturbation, i.e. the excitation spectrum and $C$ is a normalization constant. The factor 2 accounts for the fact that we have performed the integration only over positive energies. The normalization constant is obtained by calibrating the internal energy with the internal energy obtained using the temperature of on of the FD-fits in Fig. 12 at a pump-probe delay significantly exceeding the characteristic time for internal thermalization, typically at 3 ps.

In Fig. 18 where we have plotted the internal energy and the corresponding temperature obtained from the integral in eq. (13) for time delays from -1 ps to 6 ps (black circles). The light gray circles denote results from fitting a FD distribution to excitation spectra. The deviation of the energies obtained from the FD-fit and the numerical integration of spectra clearly illustrates the contribution of non-equilibrium, i.e. not-thermalized, charge carriers to the internal energy which is significant in particular at small pump-probe time-delay below one ps.

The rate $H(T_e, T_l)$ at which energy is transferred from the electronic system to the lattice can now readily be obtained by differentiation of the data in Fig. 18 with respect to time. The results of such a differentiation are reproduced in Fig. 19 where we have plotted the rate $H(T_e, T_l)$ as a function of the temperature difference $\Delta T = (T_e - T_l)$ where the lattice temperature $T_l$ is assumed to remain constant at 300 K. Similar results were obtained for a sample temperature of 40 K. Fig. 19 shows the strongly nonlinear increase of the coupling term $H$ with increasing $\Delta T$. For comparison we have included the calculated linear $\Delta T$ dependence of $H$ for copper. The latter has previously been found to have a constant slope $g$ of $10^{17}$ W/m$^3$K [61,62].

The theory of electron to lattice energy transfer by Allen [63] allows to relate the energy transfer between electrons and lattice to the electron-phonon mass enhancement parameter $\lambda$. This is done using the Eliashberg coupling function $\boldsymbol{a}^2F(\boldsymbol{w},\boldsymbol{k})$ which gives the coupling of electrons with momentum $\boldsymbol{k}$ to all final states – irrespective of their momentum – which differ from the initial state $\boldsymbol{k}$ in energy by $h\boldsymbol{w}$. The electron-phonon mass enhancement parameter is of particular interest because it allows to predict the superconducting transition temperature in conventional (BCS) superconductors using McMillans formulae [64]. According to Allen one expects a linear temperature dependence of the coupling term $H(T_e, T_l) = g(T_e - T_l)$ if the lattice temperature is similar to or above the Debye temperature $\Theta_D$ and if the temperature jump of the electronic system is significantly smaller than $T_l$. The proportionality constant then becomes $g = (3h\lambda\langle\omega^2\rangle C_e)/(\pi k_B T_l)$. This is generally appropriate for *e-ph* interactions in noble metals at room temperature, but it is not a good approximation for graphitic materials with much higher Debye temperatures. Here, we therefore use the high temperature limit of the equations derived by Allen. In this case one has a $T_e^4$ power law behavior for the couping term $H$:

$$H(T_e, T_l) = h\frac{(T_e - T_l)^5}{T_e} \tag{14}$$

with

$$h = \frac{144\boldsymbol{x}(5)k_B C_e}{\boldsymbol{p}\hbar}\frac{\boldsymbol{I}}{\Theta_D^2} \tag{15}$$

where $\boldsymbol{z}(5)$=1.0369... is Riemanns Zeta function.

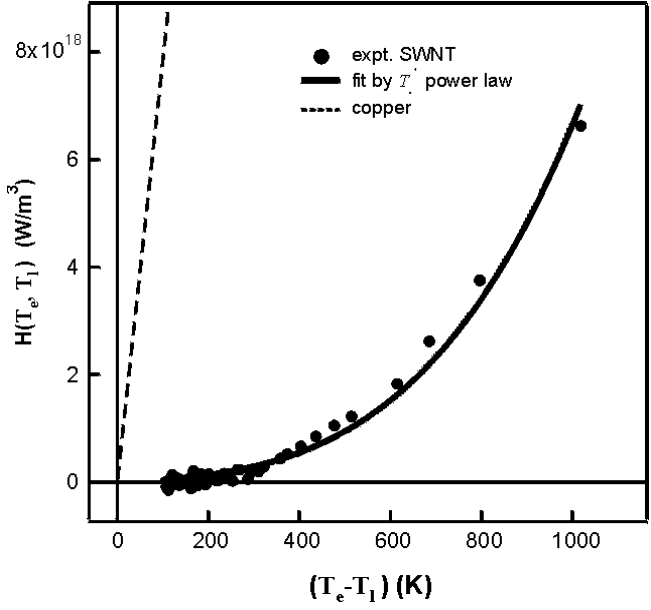

**Fig. 19.** Rate of energy transfer between electrons and lattice in SWNT bundles as a function of the temperature difference between lattice and electronic system.

The best fit to the data in Fig. 19 yields a value $\boldsymbol{1/\Theta_D^2}$ of $(6\pm2)$ $10^{-9}$K$^{-2}$. For a Debye temperature of about 1000 K [59] this gives $\boldsymbol{I}$=0.006±0.002 which is over an order of magnitude smaller than typical values obtained for the noble metals for example ($\boldsymbol{I}_{Cu}$=0.08, ref. [62]). On the other hand this value is nearly identical to that found in a similar study for graphite [65]. The major uncertainty here is the value of the on-tube Debye temperature which has been reported to be 960 K [60]. Calculated values of the on-tube Debye temperature however may be as high as 1200 K which would lead to a 40% larger value for the mass enhancement parameter.

For a temperature difference of only 100 K between electronic system and lattice these results would imply that the energy transfer rate $H$ in copper is nearly three orders of magnitude higher than that in SWNT bundles of only $5 \cdot 10^{16}$W/m$^3$. This may contribute to the high resilience of carbon nanotubes to current induced electric breakdown.

The latter can frequently be attributed to thermally as well as current activated electromigration.

In the following we will make use of the scattering formulae given by Allen and combine these with the electron-phonon mass enhancement parameter measured here to calculate elementary *e-ph* scattering times. Using the low temperature limit for the *e-ph* scattering time at the Fermi level as given by Allen [63,66]

$$\frac{1}{t_{e-ph}} = 24\boldsymbol{px}(3)\frac{\boldsymbol{I}}{1+\boldsymbol{I}}\left(\frac{T_e}{\Theta_D}\right)^3\boldsymbol{w}_D \tag{16}$$

where $\boldsymbol{z(3)}$=1.2020... and $\boldsymbol{w}_D$ is the Debye frequency, we obtain a room temperature scattering time $\tau_{e-ph}$ of 1.8 ps. Using a group velocity of $10^6$ m/s the latter would correspond to a room temperature electron-phonon mean free path of 1.8 μm. In fact, this scattering time can be attributed exclusively to scattering in metallic tube species of the SWNT bundles since the electronic heat capacity – and thus the measured cooling process – is dominated by species with high density of states at the Fermi level. Fortunately the scattering time $\tau_{e-ph}$, does not depend on the actual value of $\boldsymbol{Q}_D$ or on the value of $C_e$ as these parameters cancel out in the calculation. The only uncertainty that remains is the day to day scatter of about 30% as well as any error in the determination of the equilibrium temperature (at 3 ps pump-probe delay) which is estimated to be somewhat smaller than day to day variations.

In a previous publication [45] we used an alternative approach to demonstrate how the *e-ph* coupling strength can also be related to detailed tight-binding matrix elements of Jishi *et al.* [67]. The matrix elements used in ref. [67] were based on a calculation of the in-plane resistivity of graphene by Pietronero *et al.* [68]. Unfortunately however, the latter give an in-plane resistivity which is substantially smaller than experimental values. The mean free path and *e-ph* scattering time obtained in ref. [45] are thus somewhat larger and should be treated with caution.

Similar results for *e-ph* scattering lengths and times in SWNT bundles were obtained in conventional transport studies by Appenzeller *et al.* [69,70]. These authors reported on a *e-ph* mean free path of 11μm at a somewhat lower temperature of 250 K.

## 4 Summary and conclusions

We reported on a time-domain study of charge carrier dynamics in SWNT bundles. The experiments allow to distinguish between ultrafast *e-e* scattering processes leading to internal thermalization of an optically excited electron gas and the much slower *e-ph* interactions leading to a cooling of the heated electronic system back to the lattice temperature on the picosecond time-scale.

Electron-electron scattering times could be determined unambiguously for electron energies ranging from about 0.3 eV up to 2.3 eV above $E_F$. For states at lower energies *e-ph* interactions in addition to a refilling of states by secondary electron cascades make the determination of *e-e* scattering times more difficult. The measured lifetimes are very similar to those observed in graphite. The experiments presented here cannot discriminate between the behavior of individual tube types due to averaging by the photo-emission signal over contributions from all tubes in the SWNT bundles. The above discussion, however, may allow to predict some of the details of QP relaxation processes embedded in the averaged lifetimes. Their energy dependence, for example, can be discussed qualitatively using phase space arguments for electron relaxation in combination with $e$-$h$ pair excitation. The constraints imposed on $e$-$e$ scattering processes by conservation of energy, angular momentum, and linear momentum (along the tube/bundle axis) are expected to lead to a strong dependence of carrier lifetimes on sub-band index and energy. We predict that these constraints may occasionally also lead to an unusual energy dependence of quasiparticle lifetimes which – in some situations – can decrease with increasing energy, in strong contrast to the general behavior of a free electron system. This may also explain the extraordinary small width of the energy range facilitating resonantly enhanced Raman scattering from SWNTs [50].

The cooling of the laser heated electron gas and the corresponding rate of energy transferred to the lattice was also studied. The latter allows to determine the electron-phonon mass enhancement parameter $l$ which was found to be $0.006\pm0.002$. This coupling constant and the associated $e$-$ph$ room temperature scattering time of 1.8 ps can be associated with the metallic fraction of SWNT in the bundles because these dominate the electronic heat capacity.

In conclusion, we have presented a time-domain study of charge carrier dynamics in SWNT bundles that provides a wealth of information on $e$-$e$ and $e$-$ph$ interactions in these samples. Future studies of the temperature dependence in particular on doped samples or samples of different tube diameter may provide further insight into the details of such elementary scattering processes.

*Acknowledgements*. We thank M. Thiede for his assistance with UV-vis absorption measurements. We also would like to thank A. Rubio for valuable discussion and his comments regarding this manuscript. We furthermore thank C. Gahl and M. Wolf for stimulating discussions. R.F. acknowledges financial support by the Alexander von Humboldt foundation through a Humboldt Fellowship. It is our pleasure to acknowledge generous support by G. Ertl.

## References

1    S.J. Tans, A.R.M. Verschueren, C. Dekker, Nature **393**, 49 (1998)
2    R. Martel, T. Schmidt, H. R. Shea, T. Hertel, and Ph. Avouris, Appl. Phys. A **73**, 2447 (1998)
3    S.J. Tans, M.H. Devoret, H.J. Dai, A. Thess, R.E. Smalley, L.J. Geerligs, C. Dekker, Nature **386**, 474 (1997)
4    M. Bockrath, D.H. Cobden, P.L. McEuen, N.G. Chopra, A. Zettl, A. Thess, R.E. Smalley, Science **275**, 1922 (1997)
5    W.A. de Heer, A. Chatelain, D. Ugarte, Science **270**, 1179 (1995)
6    W.B. Choi, D.S. Chung, J.H. Kang, H.Y. Kim, Y.W. Jin, I.T. Han, Y.H. Lee, J.E. Jung, N.S. Lee, G.S. Park, J.M. Kim, Appl. Phys. Lett. **75**, 3129 (1999)
7    M.F. Yu, O. Lourie, M.J. Dyer, K. Moloni, T.F. Kelly, R.S. Ruoff, Science **287**, 636 (2000)
8    B.I. Yakobson, M.P. Campbell, C.J. Brabec,and J. Bernholc, Comp. Mat. Sci. **8**, 341 (1997)
9    J. Kong, N.R. Franklin, C.W. Zhou, M.G. Chapline, S. Peng, K.J. Cho, H.J. Dai, Science **287**, 622 (2000)
10   P.G. Collins, K. Bradley, M. Ishigami, A. Zettl, Science **287**, 1801 (2000)
11   C.M. Niu, E.K. Sichel, R. Hoch, D. Moy, H. Tennent, Appl. Phys. Lett. **70**,1480 (1997)
12   A. Rochefort, M. Di Ventra, Ph. Avouris, Appl. Phys. Lett. **78**, 2521 (2001)
13   H. Dai, E.W. Wong, and C.M. Lieber, Science **272**, 523 (1996)
14   S. Frank, Ph. Poncharal, Z. L. Wang, W.A. de Heer, Science **280**, 1744 (1998)
15   Ph. Avouris, T. Hertel, R. Martel, T. Schmidt, H.R. Shea, R.E. Walkup, Appl. Surf. Sci. **141**, 201 (1999)
16   H.T. Soh, C.F. Quate, A.F. Morpurgo, C.M. Marcus, J. Kong, and H. Dai, Appl. Phys. Lett. **75**, 627 (1999)
17   Z. Yao, C.L. Kane, and C. Dekker, Phys. Rev. Lett. **84**, 2941 (2000)
18   P.C. Collins, M.S. Arnold, and Ph. Avouris, Science **292**, 709 (2001); P.C. Collins, M. Hersam, M.S. Arnold, R. Martel, and Ph. Avouris, Phys. Rev. Lett. **86**, 3128 (2001)
19   W. Steinmann, Appl. Phys. A **49**, 365 (1989)
20   T. Fauster, Prog. Surf. Sci. **46**, 177 (1994)
21   R.W. Schoenlein, J.G. Fujimoto, G.L. Eesley and T.W. Capehart, Phys. Rev. Lett. **61**, 2596 (1988)
22   W.S. Fann, R. Storz, H.W.K. Tom and J. Bokor, Phys. Rev. Lett. **68**, 2834 (1992)
23   R. Haight, Surf. Sci. Rep. **21**, 277 (1995)
24   T. Hertel, E. Knoesel, M. Wolf and G. Ertl, Phys. Rev. Lett. **76**, 535 (1996)
25   H. Petek and S. Ogawa, Prog- Surf. Sci. **56**, 239 (1997)
26   U. Höfer, I.L. Shumay, C. Reuss, U. Thomann, W. Wallauer and T. Fauster, Chem. Phys. **251**, 111 (2000)
27   N.H. Ge, C.M. Wong, R.L. Lingle, J.D. McNeill, K.J. Gaffney and C.B. Harris, Science **279**, 202 (1998)
28   U. Höfer, I.L. Shumay, C. Reuss, U. Thomann, W. Wallauer and T. Fauster, Science **277**, 1480 (1997)
29   S.D. Brorson, A. Kazeroonian, J.S. Moodera, D.W: Face, T.K.Cheng, E.P Ippen, M.S. Dresselhaus, G. Dresselhaus, Phys. Rev. Lett. **64**, 2172 (1990)
30   A.G. Rinzler, J. Liu, H. Dai, P. Nikolaev, C.B. Huffman, F.J. Rodríguez-Macías, P.J. Boul, A.H. Lu, D. Heymann, D.T. Colbert, R.S. Lee, J.E. Fischer, A.M. Rao, P.C. Eklund, R.E. Smalley, Appl. Phys. A **67**, 29 (1998)
31   E. Knoesel, A. Hotzel, T. Hertel, M. Wolf and G. Ertl, Surf. Sci. **368**, 76 (1996)
32   E. Knoesel, A. Hotzel, and M. Wolf, Phys. Rev. B **57**, 12812 (1998)
33   R. Saito, H. Kataura. Top. Appl. Phys. **80**, 213 (2001)
34   T. Pichler, M. Knupfer, M.S. Golden and J. Fink, Phys. Rev. Lett, **80**, 4729 (1998); M. Knupfer, T. Pichler, M.S. Golden, J. Fink, A. Rinzler and R.E. Smalley, Carbon **37**, 733 (1999)
35   R. Saito, M. Fujita, G. Dresselhaus, M.S. Dresselhaus, Phys. Rev. B **46**, 1804 (1992)
36   J. Daniels, C. von Festenberg, H. Raether, and K. Zeppenfeld, Springer Tracts in Modern Physics **54**, 126 (1970); R. Ahuja, S. Auluck, J.M. Wills, M. Alouani, B. Johansson and O. Eriksson, Phys. Rev. B **55**, 4999 (1997)
37   G. Dresselhaus, M.A. Pimenta, R. Saito, J.C. Charlier, S.D.M.Brown *et al.*, Science and Application of Nanotubes, Eds. D. Tomanek and R.J. Enbody, Kluwer Academic, New York, 1999, p275-295.
38   S. Bandow, S. Asaka, Y. Saito, A.M. Rao, L. Grigorian *et al.*, Phys. Rev. Lett. **80**, 3779 (1998)
39   O. Jost, A. Gorbunov, W. Pompe, T. Pichler, R. Friedlein, M. Knupfer *et al.*, Appl. Phys. Lett. **75**, 2217 (1999)
40   J.M. Cowley, P. Nikolaev, A. Thess and R.E. Smalley, Chem. Phys. Lett. **265**, 379 (1997)
41   M.F. Lin, F.L. Shyu and R-B. Chen, Phys. Rev. B **61**, 14114 (2000); M.F. Lin, Phys. Rev. B **61**, 13153 (2000)
42   M. Damnjanovic, I. Milosevic, T. Vukovic and R. Sredanovic, Phys. Rev. B **60**, 2728 (1999)
43   R.J. Meyer, E.H.E. Pietsch, A. Kotowski (Eds.). Gmelins Handbuch der Anorganischen Chemie (Kohlenstoff.,Teil B,Lieferung 2).VCH,Weinheim,1968, 8.Au .
44   V.E. Gusev and O.B. Wright, Phys. Rev. B **54**, 2878 (1998)

45  G. Moos, C. Gahl, R. Fasel, M. Wolf and T. Hertel, Phys. Rev. Lett. **87**, 267402 (2001)

46  C.D. Spataru, M.A. Cazalilla, A. Rubio, L.X. Benedict, P.M. Echenique and S.G. Louie, Phys. Rev. Lett. **87**, 246405 (2001)

47  T. Hertel and G. Moos, Chem. Phys. Lett. **320**, 359 (2000)

48  P. Kim, T.W. Odom, J.-L. Huang, and C.M. Lieber, Phys. Rev. Lett. **82**, 1225 (1999)

49  L. C. Venema, J. W. Janssen, M. R. Buitelaar, J. W. G. Wildöer, S. G. Lemay, L. P. Kouwenhoven, and C. Dekker, Phys. Rev. B **62**, 5328 (2000)

50  A. Jorio, A.G. Souza Filho, G. Dresselhaus, M.S. Dresselhaus, R. Saito, J.H. Hafner, C.M. Lieber, F.M. Matinaga, M.S.S. Dantas, and M.A. Pimenta, Phys. Rev. B **63**, 245416 (2001)

51  E. Knoesel, A. Hotzel, T. Hertel, M. Wolf, G. Ertl, Surf. Sci. **368**, 76 (1996)

52  P.M. Echenique, J.M. Pitarke, E.V. Chulkov, A. Rubio, Chem. Phys. **251**, 1 (2000)

53  C.A. Schmuttenmaer, M. Aeschlimann, H. E. Elsayed–Ali, R. J. D. Miller, D.A. Mantell, J. Cao, Y. Gao, Phys. Rev. B. **50**, 8957 (1994)

54  J.R. Goldman, J.A. Prybyla, Phys. Rev. Lett. **72**, 1364 (1994)

55  P.G. Collins, M. Hersam, M. Arnold, R. Martel, and Ph. Avouris, Phys. Rev. Lett. **86**, 3128 (2001)

56  S.I. Anisimov, B.L. Kapeliovich, and T.L. Perel'man, Zh. Eksp. Teor. Fiz. **66**, 776 (1974) [Sov. Phys. JETP **39**, 375 (1974)]

57  J.W. McClure, I.B.M. J. Res. Develop. **8**, 255 (1964)

58  B.J.C. van der Hoeven, Jr. and P.H. Keesom, Phys. Rev. **130**, 1318 (1963)

59  J. Hone, B. Batlogg, Z. Benes, A.T. Johnson, J.E. Fischer, Science **289**, 1730 (2000)

60  J. Hone, in: Carbon Nanotubes: Synthesis, Structure, Properties and Applications, eds M.S. Dresselhaus, G. Dresselhaus, Ph. Avouris, Springer (Berlin) 2000

61  M. Bonn, D.N. Denzler, S. Funk, M. Wolf, S.-S. Wellershoff and J. Hohlfeld, Phys. Rev. B **61**, 1101 (2000)

62  S.D. Brorson, A. Kazeroonian, J.S. Moodera, D.W. Face, T.K. Cheng, E.P. Ippen, M.S. Dresselhaus, G. Dresselhaus, Phys. Rev. Lett. **64**, 2172 (1990)

63  P.B. Allen, Phys. Rev. Lett. **59**, 1460 (1987)

64  W.L. McMillan, Phys. Rev. **167**, 331 (1968)

65  G. Moos and T. Hertel, unpublished

66  P.B. Allen, Phys. Rev. **B** 11, 2693 (1975)

67  R.A. Jishi, M.S. Dresselhaus, G. Desselhaus, Phys. Rev. B **48**, 11385 (1993)

68  L. Pietronero, S. Strassler, H.R. Zeller, M.J. Rice, Phys. Rev. B **22**, 904 (1980)

69  J. Appenzeller, R. Martel, Ph. Avouris, H. Stahl, and B. Lengeler, Appl. Phys. Lett. **78**, 3313 (2001)

70  J. Appenzeller, R. Martel, and Ph. Avouris, H. Stahl, U. Th. Hunger, and B. Lengeler, Phys. Rev. B **64**, 121404 (2001)